\documentclass[11pt]{article}
\usepackage[latin1]{inputenc}
\usepackage[english]{babel}
\usepackage[namelimits]{amsmath}
\usepackage{amssymb}
\usepackage{amsmath}
\usepackage{amsthm}
\newcommand{\ac}{\'}

\begin{document}
\title{The fractal bubble model\\ 
with a cosmological constant}
\author{
Stefano Viaggiu,\\
Dipartimento di Matematica,
Universit\`a di Roma ``Tor Vergata'',\\
Via della Ricerca Scientifica, 1, I-00133 Roma, Italy.\\
E-mail: {\tt viaggiu@axp.mat.uniroma2.it}}
\date{\today}\maketitle
\centerline{To my mother Cristina Pergolini}
\begin{abstract}
We generalize the fractal bubble model (FB), recently proposed
in the literature as an alternative to the standard $\Lambda$CDM
cosmology, to include a non-zero cosmological constant.
We retain the same volume partition of voids and walls as the original
FB model, and the same matching conditions for null geodesics, but do not include
effects associated with a nonuniform time flow arising from differences of quasilocal gravitational energy
that may arise in the coarse-graining process.
The Buchert equations are written and partially integrated
and the asymptotic behaviour of the solutions is given.
For a universe with $\Lambda=0$, as it is the case in the FB model,
an initial void fraction with hyperbolic curvature evolves in such a way that
it asymptotically fills completely our particle horizon. Conversely, in presence of
a non vanishing $\Lambda$, we show that this does not happen and the voids
fill a finite fraction $f_{v_{\infty}}<1$, 
where the value of $(1-f_{v_{\infty}})$ is expected to depend on 
$\Lambda$ and the initial fraction $f_{vi}$ and also  
to be small. For its determination, a numerical integration of the equations is
necessary.
Finally, an interesting prediction of our model is a formula
giving a minimum allowed value of present day dark energy as a function
of the age of the universe and of the matter and curvature density parameters 
at our time.
\end{abstract}

\section{Introduction}
All the observations of the past decade (see in particular \cite{1,q,2}) 
are in agreement with an homogeneous and isotropic universe on large 
scales which seems to be currently in accelerating expansion.
In the standard picture based on the Friedmann-Lema\^{i}tre models (FRLW),
an accelerating universe invokes the presence of the so called dark energy.
In the FRLW picture, this dark energy is given in terms of the cosmological
constant $\Lambda$, leading to the standard cosmological model, i.e. the $\Lambda$CDM one.
As a result, about $70\%$ of the present universe would be composed of an undetectable 
non-local vacuum energy.
This dark energy represents a puzzle and perhaps the biggest problem
in modern cosmology. In fact, a direct detection of a cosmological constant is
still lacking. 
However, the dark energy is only the outcome of the Friedmann assumptions
(homogeneity and isotropy) justified by the Copernican principle.
In this respect, the observed inhomogeneities at scales $< 200$ Mpc give us the 
possibility to explore the role of inhomogeneity in cosmological models. 
In the last decade many attempts have been made (see 
for example \cite{To}-\cite{f2}) to build physically sensible inhomogeneous models.

One proposed approach 
(see for example \cite{To,13,123,13BB,16,20BB,20}) 
showed that large inhomogeneities can mimic an accelerating universe by using 
Lema\^{i}tre-Tolman-Bondi (LTB) spacetimes \cite{21,22},
although only under successful completion of several conditions \cite{3}.
In this case, the cosmological constant can be mimicked only by supposing that a huge 
spherically symmetric giant void (or hump, see \cite{z1}) of the order
of some Gpc is present, a rather unlikely assumption which is not in agreement with the CMB data
(see also \cite{z2}).

A different approach
is averaging the spatial inhomogeneities (see \cite{3B}-\cite{12}). This approach fulfils the Copernican principle, although in a statistical sense. Hence, the Copernican cosmologists introduce inhomogeneities without symmetries and then try to understand the modifications to the average evolution from backreaction. So far, the backreaction from small scales could be not enough to rule out dark energy
\cite{f8,f9}, but the debate on this issue is still open.
In this context, Wiltshire \cite{10,12} recently
proposed an interesting model, the 'fractal bubble' (FB) or 'timescape' cosmology \cite{z3},
based on the Buchert average scheme \cite{5} with two-scales mimicking the voids and walls in the 
observed web 
structure of our present day universe.
The expansion history, usually attributed to a cosmological constant, is explained by a new physical effect:
the different rate of clocks located in typical galaxies which are not expanding where the metric is
approximately spatially flat, as compared to the clocks in expanding voids where spatial curvature is 
assumed to be negative. 
A uniform Hubble flow is imposed to solve the so called Sandage-de Vaucouleurs paradox.
It is the introduction of this gauge that allows clock effects.
According with the estimates of
\cite{10,12},
clocks of isotropic observers at a volume-average location in voids presently run at  
$38\%$ faster than the ones in walls. While this result is perhaps questionable,
it is justified by Wiltshire as being the cumulative effect of a relative regional volume 
deceleration which is calculated \cite{z4} to be typically of order $10^{-10}ms^{-2}$
for most of the life of the universe, well within the expectations of the weak field regime.

In any case, although the cosmological constant remains a mystery for fundamental physics,
the
standard $\Lambda$CDM model remains the most convincing cosmological model, with much astrophysical
evidence in its support.
The standard cosmological model relies on a
universe homogeneous at any scale for any time $t$, a rather
unrealistic assumption.\\ 
As a consequence of the reasonings above, it is interesting to study 
the effects of the observed inhomogeneous structure of the universe in presence of $\Lambda$,
as for example the expansion of voids. The standard picture is to use 
perturbation theory to the standard $\Lambda$CDM model. To build an
inhomogeneous cosmological model with $\Lambda$ and without perturbation theory we use the Buchert formalism.
Within this formalism, we can gain advantage from 
the two-scale model obtained by Wiltshire, composed of walls and voids.
However, remember that the model \cite{10,12} has been built to explain 
dark energy without $\Lambda$. The observed broadly uniform Hubble flow justified in \cite{10,12} the use of the uniform Hubble gauge. It is this gauge that allows 
a non-uniform time flow. In \cite{10,12}, the non-uniform time flow is described by the 
phenomenological lapse function 
${\gamma}$. However, as we show in this paper, we can build our inhomogeneous model with $\Lambda\neq 0$
by taking the assumption of a uniform time flow (rather than a uniform Hubble flow) simply by 
changing in \cite{10,12} the interpretation of $\gamma$, which in our context is no longer
a phenomenological lapse function. It is important to stress again that the clock effects depicted in 
\cite{10,12} are only the outcome of the uniform Hubble flow used. However, note that no 
cosmological evidence of a non-uniform time flow has been observed.\\
In this paper we use the mathematical structure of
\cite{10,12}, but with $\Lambda\neq 0$ in the Buchert equations and with a phenomenological
lapse function set to unity. 
In this way, we have at our disposal a non perturbative model that will permit us 
to estimate, for example, the fraction of dark energy due to the observed inhomogeneities of our visible
universe. Finally, we can study the evolution of voids in presence of $\Lambda$ and compare this with future 
cosmological data and with the same evolution obtained with models without $\Lambda$.

In section 2 we present the model by writing, step by step, the Buchert equations in a 
workable form. In section 3 we analyze the solutions of our model.
Section 4 collects some final remarks and conclusions.
Appendix is devoted to the study of the distance-redshift relation
and to the matching conditions underlying our model.

\section{The model}

In this section we generalize step by step the calculations made in \cite{10,12} by introducing a non-vanishing
cosmological constant together with a uniform time flow gauge 
rather than the Wiltshire uniform Hubble flow gauge. As a result, clock effects disappear.  
Our aim is
to build an inhomogeneous model within the cosmological constant paradigm which includes a description
of the observed web structure of our visible universe and preserves the universal time flow of the 
concordance $\Lambda$CDM model.
We want to mimic the observed web structure of the universe by 
means of a two-scale model.

\subsection{Buchert equations for the two-scale model}
As a first assumption,
we suppose the existence of
an almost uniform time flow. Hence, all the observers in our whole particle horizon, independently on
their spatial location, use for time measurement the same coordinate $t$.
In the Buchert formalism,
we can introduce Gaussian coordinates $(t,x^i)$
which are comoving with the fluid that is assumed to be
dust filled and irrotational. 
The volume average is given to be our present particle horizon. 
For any scalar quantity $\psi(t, x^i)$, the average with respect to the volume horizon $\mathcal{H}$ is:
\begin{equation}
{<\psi(t,x^i)>}_{\mathcal{H}}=\frac{1}{V_{\mathcal{H}}}\int_{\mathcal{H}}\psi(t,x^i)\sqrt{g^{(3)}}d^3x,
\label{1}
\end{equation}
where $g^{(3)}$ denotes the determinant of the three metric on the slice a $t=const.$ and
\begin{equation}
V_{\mathcal{H}}=\int_{\mathcal{H}}\sqrt{g^{(3)}}d^3x.
\label{2}
\end{equation}
The dimensionless effective scale factor $a_{\mathcal{H}}(t)$ is given by
\begin{equation}
a_{\mathcal{H}}(t)={\left(\frac{V_{\mathcal{H}}(t)}{V_{\mathcal{H}}(t_0)}\right)}^{\frac{1}{3}}.
\label{3}
\end{equation}
The averaged expansion rate ${<\theta>}_{\mathcal{H}}$ is defined by:
\begin{equation}
{<\theta>}_{\mathcal{H}}=\frac{{\dot{V}}_{\mathcal{H}}}{V_{\mathcal{H}}}=
3\frac{{\dot{a}}_{\mathcal{H}}}{a_{\mathcal{H}}},
\label{4}
\end{equation} 
where dot indicates the time derivative. For the Hubble flow we have
$H=\frac{{<\theta>}_{\mathcal{H}}}{3}$. 
In what follows we drop the subscript
$\mathcal{H}$. The relevant exact Buchert equations (see \cite{5}) are (we use units with $c=1$):
\begin{eqnarray}
& &3\frac{{\dot{a}}^2}{a^2}=8\pi G<\rho>-\frac{\mathcal{Q}}{2}+\Lambda-
\frac{1}{2}<\mathcal{R}>,\label{5}\\
& &3\frac{\ddot{a}}{a}=-4\pi G<\rho>+\mathcal{Q}+\Lambda,\label{6}\\
& &\dot{<\rho>}=-3\frac{\dot{a}}{a}<\rho>, \label{7}\\
& &\mathcal{Q}=\frac{2}{3}\left[<{\theta}^2>-{<\theta>}^2\right]-2{<\sigma>}^2,\label{8}
\end{eqnarray}
where $\mathcal{Q}$ is the kinematical backreaction, $\sigma$ represents the shear
and $\mathcal{R}$ is the curvature of the hypersurface $t=constant$.
For the partitioning we have in mind, which is similar to the FB one, if we average over 
sufficiently large regions (see \cite{z4}), then the averaged shear $<\sigma>$ can be neglected.
The integrability condition for the system (\ref{5})-(\ref{8}) is given by
\begin{equation}
\left(6\mathcal{Q}+2<\mathcal{R}>\right)\dot{a}+a\left[\dot{\mathcal{Q}}+
<\dot{\mathcal{R}}>\right]=0.
\label{9}
\end{equation}
Equation (\ref{5}), in terms of the density parameters, can be put in the form (the so called cosmic quartet)
\begin{eqnarray}
& &{\Omega}_m+{\Omega}_{\Lambda}+{\Omega}_k+{\Omega}_{\mathcal{Q}}=1,\label{10}\\
& &{\Omega}_m=\frac{8\pi<\rho>}{3H^2},\;{\Omega}_k=-\frac{<\mathcal{R}>}{6H^2},\;
{\Omega}_{\mathcal{Q}}=-\frac{\mathcal{Q}}{6H^2},\;
{\Omega}_{\Lambda}=\frac{\Lambda}{3H^2},\label{11}
\end{eqnarray}
where $H$ is the Hubble rate function.

For the partitioning of our visible universe, we follow the two-scales model of the 
original FB model (see \cite{10,12} for more). First of all we must specify the so called 
finite infinity regions. 
In the universe, there exist regions expanding and contracting. However, there are regions, called
finite infinity $F_I$ (see \cite{12,f4}), that are the set of timelike boundaries of compact 
{\bf disjoint} domains
$I$
for which the average expansion is zero and becomes positive outside, i.e. for any surface $t=const$:
\begin{eqnarray}
& &{<\theta>}_{{\bigcup}_{I}F_I}=0,\label{t1}\\
& &\exists\;D_I,\;\;F_I\supset D_I\;\rightarrow\;{<\theta>}_{{\bigcup}_I D_I}>0.
\nonumber
\end{eqnarray}
In other words, ${\bigcup}_I F_I$  contains regions which on average are non-expanding.
Our position is within a $F_I$ region. 
The $F_I$ regions 
are well within the wall regions, which are defined as the disjoint regions
that have on average a spatially flat metric with expansion factor $a_w(t)$:
\begin{equation}
ds^2_w=-dt^2+a^2_w\left[d{\eta}^2_w+{\eta}^2_wd{\Omega}^2\right].
\label{t2}
\end{equation}
Outside the wall regions (the region complementary to the walls with respect to
the particle horizon) we have the void ones that are defined as expanding
regions where the metric is on average hyperbolic with scale factor $a_v(t)$
\begin{equation}
ds^2_v=-dt^2+a^2_v\left[d{\eta}^2_v+{\sinh}^2({\eta}_v)d{\Omega}^2\right].
\label{t3}
\end{equation}
Another crucial assumption we make (also in the FB model) is the 
presence of a scale of homogeneity. This assumption is justified by the fact that statistical analysis 
of the galaxy distributions from available data seem to indicate an homogenization on scales of order
of $(100-200)Mpc$. 
By scale of homogeneity we mean
that averages of any
variable beyond this scale practically will no longer depend on the scale.
For the Hubble function in walls and voids we have:
\begin{equation}
H_w=\frac{{<\theta>}_w}{3}=\frac{{\dot{a}}_w}{a_w},\;
H_v=\frac{{<\theta>}_v}{3}=\frac{{\dot{a}}_v}{a_v}.
\label{12}
\end{equation}
With '$a$' we denote the scale factor at the scale of homogeneity.
The Buchert equations for $a(t)$ are given by (\ref{5})-(\ref{8}). 
Moreover, in what follows, we denote with the subscript 'i' an initial early time $t_i$.
Since the main cosmological data (distance-redshift relation)
are available starting from the recombination era, we intend with $t_i$ this
early time. Furthermore, we denote with $V_i$ the initial volume of the particle horizon at the 
recombination era.
For the average on the volume representing the whole 
visible universe (particle horizon) defined by $V=a^3 V_i$ we have
\begin{equation}
a^3=f_{wi}\;a_{w}^3+f_{vi}\;a_{v}^3,\;\;\;f_{wi}+f_{vi}=1,
\label{13}
\end{equation}
where $f_{wi}$ and $f_{vi}$ are initial fractions of walls and voids. 
Formula (\ref{13}) relates the expansion factors at the two-scales (walls and voids)
to the average geometry represented by $a(t)$ at the scale of homogeneity.
The wall volume is $V_w=V_{wi}a_{w}^3$ while the void volume is
$V_v=V_{vi}a_{v}^3$, being $f_{wi}=\frac{V_{wi}}{V_i}$ and
$f_{vi}=\frac{V_{vi}}{V_i}$. For a time $t>t_i$ the picture is
\begin{equation}
f_v(t)+f_w(t)=1,\;\;f_w=f_{wi}\frac{a_{w}^3}{a^3},\;\;f_v=f_{vi}\frac{a_{v}^3}{a^3}.
\label{14}
\end{equation}
At the volume horizon, for the Hubble rate we have
\begin{equation}
H=f_w H_w+f_v H_v.
\label{15}
\end{equation}
Moreover, we define $I_w$ and $I_v$ according to
\begin{equation}
H=I_w H_w=I_v H_v.
\label{16}
\end{equation}
In the Wiltshire model we have, instead of $I_w$, the 
phenomenological lapse function $\gamma$.
Differently to the FB model, in our model $I_w, I_v$ are not related to possible time delay
effects. In practice, we adopt the same partitioning present in \cite{12}, but in a different context,
i.e. with a uniform time flow gauge instead of a uniform Hubble one.
As a consequence, in our context $I_w$ is only a measure of the ratio between the averaged Hubble flow
at the scale of homogeneity and the one at the wall scale. This choice can be justified by the fact that,
at present day, no evidence for a non-uniform time flow at cosmological scales has been observed,
although the converse cannot be excluded.
By derivating $f_v(t)$ given by (\ref{14}) we have:
\begin{equation}
{\dot{f}}_v=-{\dot{f}}_w=3(1-f_v)(1-I_w^{-1})H=
\frac{3f_v(1-f_v)(1-h)H}{h+(1-h)f_v},
\label{17}
\end{equation}
where $h$ is defined by $I_v=h I_w$. For a general scalar $\psi$ we have:
\begin{equation}
<\psi>=f_w{\psi}_w+f_v{\psi}_v.
\label{18}
\end{equation}
As a consequence of (\ref{18}) we get:
\begin{equation}
<{\theta}^2>=9(f_w H_w^{2}+f_v H_v^{2}).
\label{19}
\end{equation}
From (\ref{8}),(\ref{15}), (\ref{17}) and (\ref{19}) and neglecting the possibility of intrinsic
backreaction within the walls and voids (see\cite{7,z5}) we obtain:
\begin{equation}
\mathcal{Q}=6f_v(1-f_v){(H_v-H_w)}^2=
\frac{6f_v(1-f_v){(1-h)}^2H}{{[h+(1-h)f_v]}^2}.
\label{20}
\end{equation}
A useful alternative expression for the back-reaction
$\mathcal{Q}$ is:
\begin{equation}
\mathcal{Q}=\frac{2{{\dot{f}}_v}^2}{3f_v(1-f_v)}.
\label{21}
\end{equation}
Concerning the mean three curvature $<\mathcal{R}>$,  
we have ${<\mathcal{R}>}_v=6\frac{k_v}{a_v^2}$
and ${<\mathcal{R}>}_w=6\frac{k_w}{a_w^2}=0$ since by 
construction $k_w=0$. Hence, by setting
$q^2=-k_v {f_{vi}}^{\frac{2}{3}}$, thanks to 
(\ref{18}),
we get:
\begin{equation}
<\mathcal{R}>=-\frac{q^2 f_{v}^{\frac{1}{3}}}{a^2}.
\label{22}
\end{equation}
Equation (\ref{7}) can be easily integrated obtaining
$<\rho>={\rho}_0 a_0^3/a^3$. Finally, the independent equations
(\ref{5}) and (\ref{9}) become, with ${\dot{f}}_v\neq 0$
\begin{eqnarray}
& &\frac{{\dot{a}}^2}{a^2}+\frac{{{{\dot{f}}_v}}^2}{9f_v(1-f_v)}-
\frac{q^2 {f_v}^{\frac{1}{3}}}{a^2}=
\frac{8}{3}\pi G{\rho}_0\frac{a_0^3}{a^3}+\frac{\Lambda}{3},
\label{23}\\
& &{\ddot{f}}_v+\frac{{{\dot{f}}_v}^2(2f_v-1)}{2f_v(1-f_v)}+
3\frac{\dot{a}}{a}{\dot{f}}_v-
\frac{3q^2{{f_v}}^{\frac{1}{3}}(1-f_v)}{2a^2}=0.
\label{24}
\end{eqnarray}
Equations (\ref{22}) and (\ref{23}) are the master equations of our model involving
the scale factor $a(t)$ at the scale of homogeneity and the void fraction $f_v(t)$.
In the following subsection we obtain from (\ref{23})-(\ref{24}) a function which allows
to reduce the Buchert equations in a workable form. 
     
\subsection{A useful function for the model}
First of all,
equation (\ref{6}) can be written in the following way
\begin{equation}
\frac{\ddot{a}}{a}=-\frac{4\pi G}{3}{\rho}_0\frac{a_0^3}{a^3}+
\frac{2}{9}\frac{{\dot{f}}_v^2}{f_v(1-f_v)}+\frac{\Lambda}{3}.
\label{32}
\end{equation}
By considering equations (\ref{23}), (\ref{24}) and (\ref{32}) we obtain
\begin{equation}
6\frac{\ddot{a}}{a}+3\frac{{\dot{a}}^2}{a^2}-2\frac{{\ddot{f}}_v}{1-f_v}-6
\frac{\dot{a}}{a}\frac{{\dot{f}}_v}{(1-f_v)}-\frac{{\dot{f}}_v^2}{{(1-f_v)}^2}=
3\Lambda.
\label{33}
\end{equation}
From equation (\ref{17}) we get
\begin{equation}
I_w=\frac{3(1-f_v)\dot{a}}{3(1-f_v)\dot{a}-{\dot{f}}_v a}.
\label{34}
\end{equation}
Moreover, we have
\begin{equation}
{\Omega}_m=\frac{8\pi G{\rho}_0 a_0^3}{3H^2 a^3},\;
{\Omega}_k=\frac{q^2 f_v^{\frac{1}{3}}}{a^2 H^2},\;
{\Omega}_{\mathcal{Q}}=\frac{-{\dot{f}}_v^2}{9f_v(1-f_v)H^2}.
\label{35}
\end{equation}
Finally, by multiplying (\ref{33}) by $24\pi G{\rho}_0 a_0^3/{[{\dot{f}}_v a-3(1-f_v)\dot{a}]}^3$ and
with the help of (\ref{35}) and (\ref{23})-(\ref{24}), we get
\begin{equation}
\frac{d}{dt}\left(\frac{I_w^2{\Omega}_m}{1-f_v}\right)=
-\frac{I_w^3{\Omega}_m}{(1-f_v)}\frac{a}{\dot{a}}\Lambda.
\label{36}
\end{equation}
After integrating and using (\ref{16}) we have:
\begin{equation}
\frac{(1-{\epsilon}_i)I_w^2{\Omega}_m}{(1-f_v)}=
{\Omega}_F=e^{-\int_{t_i}^{t}\frac{\Lambda}{H_w}dt},
\label{37}
\end{equation}
where ${\epsilon}_i$ is a small integration constant ($<<1$) which will be determined
later. 
In the original FB model, in the limit $\Lambda=0$, 
the function ${\Omega}_F$ is a constant. Its value is $1$.
In the FB model its presence is justified
in the sense that, if we consider only matter in walls, then there must exist a constant critical 
density of reference (namely ${\Omega}_F=1$). This critical parameter denotes the  
density at the recombination era within walls. After the recombination, the voids
begin to expand and occupy the particle horizon. In our $\Lambda$FB model this picture 
is no longer valid. 
The function ${\Omega}_F$ begins with the value $1$ at the recombination
and it reaches $0$ asymptotically. We have denoted this function as a density parameter
by recalling the original meaning of this function in the Wiltshire model.
However, we are not able to indicate a clear physical interpretation for
${\Omega}_F$.
In our context, formula (\ref{37}) is very useful and allows to reduce, simplify and partially integrate the 
Buchert equations, as we see in the next subsection.

\subsection{Final form of the equations}
We now write down the Buchert equations in a workable form.
Combining the equations (\ref{23}), (\ref{24}), (\ref{34}) and (\ref{37}) one obtains
\begin{equation}
\frac{{\dot{a}}^2}{a^2}(2f_v-1)+\frac{2{\dot{f}}_v\dot{a}}{3a}-\frac{q^2}{a^2}
{f}_v^{\frac{4}{3}}+\frac{8\pi G}{3}{\rho}_0\frac{a_0^3}{a^3}
\left[\frac{(1-{\epsilon}_i)}{{\Omega}_F}-f_v\right]=\frac{\Lambda}{3}f_v\;.
\label{38}
\end{equation}
The main Buchert equations become (\ref{23}) and (\ref{38}) together with the
relation (\ref{37}). We can further simplify these equations using manipulations
similar to those found in \cite{10,12}.\\
Namely, a first equation is obtained by multiplying equation (\ref{23}) by $f_v$ 
and subtracting equation (\ref{38}). A second equation is obtained by multiplying
equation (\ref{23}) by $(1-f_v)$ and adding equation (\ref{38}). As a result we get
\begin{eqnarray}
& &(1-f_v)\frac{\dot{a}}{a}-\frac{{\dot{f}}_v}{3}=
\sqrt{\frac{8}{3}\pi G{\rho}_0\frac{a_0^3}{a^3}\left(\frac{1-{\epsilon}_i}{{\Omega}_F}
\right)(1-f_v)},\label{39}\\
& &\frac{\dot{a}}{a}+\frac{{\dot{f}}_v}{3f_v}=\frac{q}{a{f}_v^{\frac{1}{3}}}
\sqrt{1+\frac{\Lambda}{3}\frac{a^2}{q^2{f}_v^{}\frac{1}{3}}
+\frac{8\pi G{\rho}_0}{3q^2{f}_v^{\frac{1}{3}}}\frac{a_0^3}{a}
\left(1+\frac{{\epsilon}_i-1}{{\Omega}_F}\right)}. \label{40}
\end{eqnarray}
Thanks to (\ref{35}), equations (\ref{39}) and (\ref{40}) become:
\begin{eqnarray}
& &(1-f_v)\frac{\dot{a}}{a}-\frac{{\dot{f}}_v}{3}=
\sqrt{{\Omega}_{0m}H_0^2\frac{a_0^3}{a^3}\left(\frac{1-{\epsilon}_i}{{\Omega}_F}
\right)(1-f_v)},\label{41}\\
& &\frac{\dot{a}}{a}+\frac{{\dot{f}}_v}{3f_v}=\frac{a_0 H_0}{a{f}_v^{\frac{1}{3}}}
\sqrt{\frac{{\Omega}_{0k}}{f_{v_0}^{\frac{1}{3}}}+
\frac{{\Omega}_{0\Lambda}a^2}{f_v^{\frac{1}{3}}a_0^2}
+\frac{a_0{\Omega}_{0m}}{a{f}_v^{\frac{1}{3}}}
\left(1+\frac{{\epsilon}_i-1}{{\Omega}_F}\right)}. \label{42}
\end{eqnarray}
A further manipulation can be done by multiplying
equation (\ref{41}) by\\ 
$H_0^{-1}{(1-f_v)}^{-\frac{2}{3}}a$ and equation (\ref{42})
by $a_0^{-1}H_0^{-1}f_v^{\frac{1}{3}}a$. We get:
\begin{eqnarray}
& &\frac{1}{H_0}\frac{d}{dt}\left[{(1-f_v)}^{\frac{1}{3}}a\right]=
\sqrt{\frac{{\Omega}_{0m}a_0^3}{{(1-f_v)}^{\frac{1}{3}}a}
\frac{(1-{\epsilon}_i)}{{\Omega}_F}},\label{43}\\
& &\frac{1}{H_0}\frac{du}{dt}=
\sqrt{\frac{{\Omega}_{0k}}{f_{v_0}^{\frac{1}{3}}}\left[
1+\frac{{\Omega}_{0\Lambda}}{{\Omega}_{0k}}
\frac{f_{v_0}^{\frac{1}{3}}}{f_v}\;u^2+
\frac{{\Omega}_{0m}f_{v_0}^{\frac{1}{3}}}{{\Omega}_{0k}\;u}
\left(1+\frac{{\epsilon}_i-1}{{\Omega}_F}\right)\right]},
\label{44}
\end{eqnarray}
where $u={f}_{vi}^{\frac{1}{3}}\frac{a_v}{a_0}$.
Equations (\ref{43})-(\ref{44}) represent the final form of our field equations.
We are now ready to analyze the main features of our model.

\section{A study of the equations}

\subsection{The case with $\Lambda=0$}
In the limit $\Lambda=0$ we formally recover the solution in \cite{10} for the Buchert 
equations (\ref{43}) and (\ref{44}), but without clock effects.
It is important to note that we do not use approximations. The absence of clock effects is only
the outcome of our chosen gauge, i.e. the uniform time flow
which is different from the one used by the original Wiltshire model. 
As a consequence, we formally obtain, in the limit $\Lambda=0$, the same master equations of
\cite{10,12} expressed in terms of the time $t$, but with a 
phenomenological lapse function set to unity (uniform time flow gauge).  
This happens because a different interpretation for the parameter $I_w$ is present in our model due to the 
different gauge used.
Hence, in this limit, the wall factor $a_w$ evolves exactly as an Einstein-de Sitter universe.
Furthermore, as it is shown in \cite{10,z3}, the general $\Lambda=0$ solution possesses a 
tracking limit in which the factor $a_v$
evolves approximatively as a Milne universe at late times.
For $f_v$ the tracking limit solution is given in \cite{10,z4}:
\begin{equation}
f_v=\frac{3f_{v_0}H_0 t}{3f_{v_0}H_0 t+(1-f_{v_0})(2+f_{v_0})}. 
\label{45}
\end{equation}
Formula (\ref{45}) shows that in absence of a cosmological constant, the voids
monotonically expand and asymptotically reach the value $1$, i.e. they occupy the full
volume horizon.
The deceleration parameter 
$q=2{(1-f_v)}^2/{(2+f_v)}^2$ is always positive and reaches asymptotically zero from positive
values.

\subsection{The case with $\Lambda\neq 0$}

\subsubsection{Late times behaviour of the solutions}
First of all, we consider the far future limit of our model. This behaviour can be exactly determined
by analyzing the field equations.
In this limit, both the wall density ${\rho}_w$ and the void density ${\rho}_v$
are negligible and asymptotically become zero. Hence, in this limit, 
both $a_w$ and
$a_v$ are driven only by the cosmological constant. As a result, for
$t\rightarrow\infty$ we must have:
\begin{equation}
a_w(t)\sim e^{\sqrt{\frac{\Lambda}{3}}t},\;\; 
a_v(t)\sim e^{\sqrt{\frac{\Lambda}{3}}t}\;\rightarrow\;
a\sim e^{\sqrt{\frac{\Lambda}{3}}t}.
\label{47}
\end{equation}
Expressions (\ref{47}) states that asymptotically, since the universe is dominated by the 
cosmological constant, both walls and voids must evolve as a de Sitter spacetime.
Thus, asymptotically we have
\begin{equation}
H_w(t\rightarrow\infty)=H_v(t\rightarrow\infty)=H(t\rightarrow\infty)=\sqrt{\frac{\Lambda}{3}}.
\label{48}
\end{equation}
As a result, by taking the asymptotic limit in (\ref{23}), we obtain
\begin{equation}
\lim_{t \rightarrow \infty} \frac{{\dot{f}}_v^2}{f_v(1-f_v)} = 0
\label{49}
\end{equation}
Hence ${\dot{f}}_v(t\rightarrow\infty)=0$ independently on the behaviour of $f_v$.
Formula (\ref{49}) implies that the backreaction $\mathcal{Q}$ reaches asymptotically zero.
At late times the Buchert equations for $a_w$ and $a_v$ decouple and the exact GR
solutions dominated by the cosmological constant arise. Note that the same phenomenon
appears in the original Wiltshire model, where at late times the solutions for $a_w$ and
$a_v$ look like the exact GR solutions and $\mathcal{Q}$ vanishes.
Moreover, from equation (\ref{37}),
for the parameter ${\Omega}_F$, at early times we have
${\Omega}_F\simeq 1$, while asymptotically ${\Omega}_F\rightarrow 0$.
The exact expression for ${\Omega}_F$ will be calculated later.

Another interesting asymptotic limit is the one involving $f_v$. Its value at early times, 
to be in agreement with WMAP data, is
fixed to be  a priori $f_{vi}<<1$. From (\ref{43}) or (\ref{44}), thanks to (\ref{47})
and (\ref{48}), for $f_{v_{\infty}}$ we have
\begin{equation}
f_{v_{\infty}}=1-\frac{{\Omega}_{0m}}{{\Omega}_{0\Lambda}}k_{\infty}(1-{\epsilon}_i),
\label{51}
\end{equation}
where:
\begin{equation}
\lim_{t \rightarrow \infty} \frac{a_0^3}{a^3{\Omega}_F} = k_{\infty}.
\label{52}
\end{equation}
Concerning $k_{\infty}$, from equation (\ref{43}) we have
\begin{equation}
\frac{1}{H_0^2}\frac{f_{wi}\;{\dot{a}}_w^2 a_w}{{\Omega}_{0m}\;a_0^3}=
\frac{(1-{\epsilon}_i)}{{\Omega}_F}.
\label{53}
\end{equation}
We multiply (\ref{53}) by $\frac{a_0^3}{a^3}$. By performing the limit
$t\rightarrow\infty$, we get
\begin{equation}
k_{\infty}=\frac{1}{(1-{\epsilon}_i)}
\lim_{t \rightarrow \infty} \frac{f_{wi}{\dot{a}}_w^2 a_w}
{H_0^2{\Omega}_{0m}\;a^3}.
\label{54}
\end{equation}
Formula (\ref{54}), thanks to (\ref{47}), implies that $k_{\infty}\neq 0$.
As a consequence, differently from the $\Lambda=0$ case, $f_{v_{\infty}}\neq 1$.
For its determination, a numerical integration of the equations is necessary.
In any case, some considerations can be made. In the limit $\Lambda=0$, we must 
have $f_{v_{\infty}}=1$ and $f_{v_{\infty}}$ must be a function of $f_{vi}$
and of the adimensional quantity $\sqrt{\Lambda}t_i$. Since $\sqrt{\Lambda}t_i<<1$, on general grounds
we expect that $(1-f_{v_{\infty}})<<1$. However, 
it is possible 
to calculate this value
only by a direct numerical computation.
We only mention this interesting difference with respect to the 
FB model.

\subsubsection{Explicit expressions for $a_w$ and ${\Omega}_F$}
In this subsection, we integrate directly our field equations obtaining the exact expressions
for $a_w$ and ${\Omega}_F$ which allow to obtain further asymptotic behaviours and confirm the ones above.  
After differentiating (\ref{53}) and applying again (\ref{53}) we get the equation:
\begin{equation}
2\frac{{\ddot{a}}_w}{a_w}+\frac{{\dot{a}}_w^2}{a_w^2}=\Lambda.
\label{55}
\end{equation}
Equation (\ref{55}) is nothing but the exact Einstein equation for $a_w$.
Hence, the wall factor $a_w$ solution of the Buchert
equations is exactly the same of the $\Lambda$CDM model. This does not represent a surprise.
In fact, in the model \cite{10,12} the wall metric evolves as the exact Einstein solution, while the void
metric evolves in a different manner with respect to the decoupled exact Einstein solution.
As a result we have
\begin{equation}
a_w= a_{w0}\;{\sinh}^{\frac{2}{3}}\left(\frac{\sqrt{3\Lambda}}{2}t\right).
\label{56}
\end{equation}
Moreover:
\begin{equation}
H_w=
\frac{\cosh\left(\sqrt{\frac{3\Lambda}{4}}\;t\right)}
{\sinh\left(\sqrt{\frac{3\Lambda}{4}}\;t\right)}\;\sqrt{\frac{\Lambda}{3}}.
\label{57}
\end{equation}
Expressions (\ref{56}) and (\ref{57}) confirm the estimation (\ref{47}) and (\ref{48}).
Furthermore, we calculate ${\Omega}_F$. We obtain
\begin{equation}
{\Omega}_F={\left(\frac{\cosh\left(\sqrt{\frac{3\Lambda}{4}}\;t_i\right)}
{\cosh\left(\sqrt{\frac{3\Lambda}{4}}\;t\right)}\right)}^2.
\label{58}
\end{equation}
Equation (\ref{43}) can be easily integrated to give
\begin{equation}
{(1-f_v)}^{\frac{1}{3}} a=
a_0{\left(\frac{{\Omega}_{0m}(1-{\epsilon}_i)}{{\Omega}_{0\Lambda}}\right)}^{\frac{1}{3}}
\frac{{\sinh}^{\frac{2}{3}}\left(\frac{\sqrt{3\Lambda}}{2}t\right)}
{{\cosh}^{\frac{2}{3}}\left(\frac{\sqrt{3\Lambda}}{2}t_i\right)}.
\label{59}
\end{equation}
Finally, for $a_{w0}$ we have
\begin{equation}
a_{w0}=\frac{a_0}{f_{wi}^{\frac{1}{3}}}
{\left(\frac{{\Omega}_{0m}(1-{\epsilon}_i)}{{\Omega}_{0\Lambda}}\right)}^{\frac{1}{3}}
\frac{1}{{\cosh}^{\frac{2}{3}}\left(\frac{\sqrt{3\Lambda}}{2}t_i\right)}.
\label{60}
\end{equation}
Equation (\ref{44}) cannot be integrated in a elementary way. 
For late times, this equation reduces to the exact Einstein
equation, i.e. without backreaction. Note that we could gain advantage from this 
fact to obtain an approximate expression for $f_v(t)$.

\subsubsection{Early and late times behaviours for $a_v$, $f_v$ and $I_w$}
Equation (\ref{44}) can be useful to calculate the early
times behaviour of our model. At early times, we have
\begin{equation}
a_w=a_{w0}{\left(\frac{3\Lambda}{4}\right)}^\frac{1}{3}t^{\frac{2}{3}},\;\;
a_v=a_{v0}\;t^{\frac{2}{3}},
\label{61}
\end{equation}
where the behaviour for $a_v$ has been obtained by evaluating (\ref{44}) at $t\simeq t_i$.
By putting again the above expression for $a_v$ in (\ref{44}), we obtain
\begin{equation}
a_{v0}=a_0{\left(\frac{9}{4}\right)}^{\frac{1}{3}}
\frac{H_0^{\frac{2}{3}}}{f_{vi}^{\frac{1}{3}}}{\left(
{\Omega}_{0m}{\epsilon}_i\right)}^{\frac{1}{3}}.
\label{62}
\end{equation}
After putting (\ref{62}) in (\ref{59}), we get the relation 
$f_{vi}={\epsilon}_i$. In this way, all the parameters of the model are
specified. With the same technique, we can obtain the asymptotic behaviour of
$a_v$. We get $a_v=a_{v_{\infty}}e^{\sqrt{\frac{\Lambda}{3}} t}$ where
\begin{equation}
a_{v_{\infty}}^3=\frac{{\Omega}_{0m}a_0^3}{{\Omega}_{0\Lambda}f_{vi}H_0^2}
\left(\frac{f_{v_{\infty}}}{1-f_{v_{\infty}}}\right).
\label{63}
\end{equation}
Concerning the asymptotic behaviour of $f_v$, from (\ref{43}) at the leading term in $t$ we have
\begin{equation}
f_v=f_{v_{\infty}}-2\frac{H_0 B}{\Lambda}\sqrt{{\Omega}_{0\Lambda}f_{v_{\infty}}}
e^{-\frac{\sqrt{3\Lambda}}{2} t}+o(1),
\label{64}
\end{equation}
being $B$ an integration constant. For $B>0$, $f_v$ reaches its asymptotic value from below,
conversely if $B<0$. Since we expect $(1-f_{v_{\infty}})<<1$, perhaps $B>0$.

As an important consequence of (\ref{49}), we can study the asymptotic behaviour of 
$I_w$. In fact, from (\ref{34}) we get
\begin{equation}
I_w=\frac{1}{1-\frac{a{\dot{f}}_v}{3(1-f_v)\dot{a}}}.
\label{50}
\end{equation}
Thus, both at early and asymptotic times $I_w=1$.
This behaviour can be understood thanks to (\ref{47}).
Furthermore, the behaviour given by (\ref{50}) can potentially account
for the broadly observed Hubble flow, without invoking clock effects.
This concludes a preliminary study of our solutions.

\subsubsection{A formula for our model and its consequences}

We derive a very useful and intriguing formula containing all the cosmological
parameters of our model. In fact, by combining the equations (\ref{41}) and (\ref{42}) and
evaluating them at a given present time $t_0$, we get
\begin{equation}
1=\sqrt{{\Omega}_{0m}(1-f_{v_0})\frac{(1-f_{vi})}{{\Omega}_{0F}}}
+\sqrt{f_{v_0}}\sqrt{{\Omega}_{0k}+{\Omega}_{0\Lambda}+
{\Omega}_{0m}\left(1+\frac{f_{vi}-1}{{\Omega}_{0F}}\right)},
\label{65}
\end{equation}
where ${\Omega}_{0F}$, given by (\ref{58}), 
is also calculated at  
$t_0$. This equation is a generalization of a similar relation present in \cite{10}.
An important consequence of (\ref{65}) is provided by its domain
\begin{equation}
{\Omega}_{0\Lambda}\geq {\Omega}_{0m}\left(-1+\frac{1-f_{vi}}{{\Omega}_{0F}}\right)-
{\Omega}_{0k}.
\label{66}
\end{equation}
The inequality (\ref{66}) provides an absolute minimum for 
${\Omega}_{0\Lambda}$ (at least for positive curvatures) 
in terms of the actual matter and 
curvature density parameters and of the
age of the universe. From (\ref{66}) and (\ref{37}), we deduce that
${\Omega}_{0\Lambda}\geq f_{vi}$ at the recombination era and reaches the 
asymptotic limit $>(1-f_{v_\infty})$ at late times i.e, thanks to (\ref{10}), 
${\Omega}_{\Lambda}$ reaches asymptotically the value $1$.\\ 
As a first consideration, (\ref{66}) implies that,
with ${\Omega}_{0m}$ and ${\Omega}_{0k}$ held fixed, a higher age of the universe requires
a higher value for ${\Omega}_{0\Lambda}$, i.e. more dark energy.
Moreover, formula (\ref{66}) also says that,
in order to mimic dark energy, a large amount of negative curvature is needed.
While a large negative curvature seems in disagreement with analysis of the CMB,
this calculation was obtained within the $\Lambda$CDM paradigm and is not directly relevant here.\\
It should be stressed that some studies
\cite{f5,f6} on the
evolution of voids fraction in the $\Lambda$CDM context estimate an actual fraction of voids
between $(40-70)\%$ of the total particle horizon volume.
In these studies voids are identified as underdense regions. However, it should be noted that
an underdense region may not have necessarily a negative curvature on average:
a sufficient amount, for example, of dark matter can be enough to make the average curvature
non-negative.\\ 
In the following numerical examples, we assume that the fraction
of voids present in the universe has negative curvature.
We calculate the cosmological parameters that allow to have
$(40-70)\%$ of voids with hyperbolic curvature.
We now give some crude numerical examples.  
As an example, we pose  
$t_0\simeq 14$ Gyr with $\Lambda\sim 10^{-35}/s^2$
and so we have ${\Omega}_{0F}\simeq 1/(3.33)$ and 
we pose ${\Omega}_{0m}=0.3$. 
Hence, from (\ref{65}),
by setting (remember that
$t_0$ and ${\Omega}_{0m}$ are held fixed)
${\Omega}_{0k}=0.1$, ${\Omega}_{0\Lambda}=0.72$, we have $f_{v_0}=0.4$ and 
${\Omega}_{0\mathcal{Q}}=-0.12$. By setting for example 
${\Omega}_{0k}=0.2$, ${\Omega}_{0\Lambda}=0.72$, ${\Omega}_{0\mathcal{Q}}=-0.22$, we
have $f_{v_0}=0.6$. Finally, by choosing
${\Omega}_{0\Lambda}=0.72$, ${\Omega}_{0k}=0.28$, ${\Omega}_{0\mathcal{Q}}=-0.3$, we obtain
$f_{v_0}=0.7$. These crude estimations show that
a large (observed) fraction of voids could be in agreement with a relatively small value for 
${\Omega}_{0k}$. Furthermore, in the examples quoted above we have
${\Omega}_{0\mathcal{Q}}+{\Omega}_{0k}\simeq 0$. As a result, we could have an inhomogeneous model
with a fraction of voids $f_{v_0}\geq 0.4$ in agreement with the one actually observed
and with the same present day values for ${\Omega}_{0\Lambda}$
and ${\Omega}_{0m}$ of the $\Lambda$CDM model. This suggests that the backreaction $\mathcal{Q}$,
rather than acting as a dark energy,
could act on void scales as an effective positive curvature, balancing the 
negative curvature of the voids themselves.
By describing a
cosmological model where
\begin{equation}
{\Omega}_{0m}+{\Omega}_{0\Lambda}\simeq 1,\;\;
{\Omega}_{0\mathcal{Q}}+{\Omega}_{0k}\simeq 0,
\label{67}
\end{equation}
and with a large fraction of voids, we can obtain a model mimicking the relation 
${\Omega}_{0m}+{\Omega}_{0\Lambda}=1$ of the flat $\Lambda$CDM one with the further constraint
${\Omega}_{0\mathcal{Q}}+{\Omega}_{0k}= 0$.

As a final consideration for this section, note that inequality (\ref{66}) can be seen also in
terms of the parameter ${\Omega}_{0\mathcal{Q}}$. Thanks to (\ref{10}) we have
\begin{equation}
{{\Omega}_{0\mathcal{Q}}}\leq 1-\frac{{\Omega}_{0m}}{{\Omega}_{0F}}(1-f_{vi})
\label{68}
\end{equation}
and from (\ref{66}) and (\ref{68}) we deduce that the 
density parameter ${\Omega}_{0m}$ plays an important role. For an overview on the backreaction 
issue, see \cite{f7,f8,f9}.

\section{Conclusions}
Many attempts have been made in the literature in order to explain or rule out the cosmological constant
in the present day cosmological models. In particular, within the Buchert formalism, the FB model 
\cite{10,12} has been recently presented to explain dark energy in terms of the gravitational 
energy stored in voids. The amount of such a gravitational energy sufficient to rule out 
the presence of $\Lambda$ is rather huge ($\simeq\;38\%$ at present day). 
Within the formalism used for 
the FB model, it is possible to build a more realistic model of universe than the ones
obtained with exact solutions as, for example, LTB metrics. In fact, to 
mimic dark energy, such exact solutions
require a single huge void or hump of order of Gpc. Such a picture is in contradiction
with the known 'geography' of the universe, where a web structure seems to be more appropriate.
An important issue is represented by the calculation of the effects of inhomogeneities on the 
value of $\Lambda$. It is very useful to build a 
non-perturbative model,
within the cosmological constant paradigm, mimicking the 
present day web structure of the universe.
Our main aim was to present a model allowing  
to study, in a non-perturbative
framework,
the effects of the observed inhomogeneities on the cosmological parameters
and on the cosmic evolution. 
We have generalized the FB model by introducing a cosmological constant
but without introducing ab initio clock effects. We have written the relevant equations in a simple way.
In the limit $\Lambda=0$ the equations reduce to the form presented in \cite{10,12}. We partially integrate
and study the asymptotic behaviour of the solutions. A first difference with respect to the 
FB model is that the cosmological constant inhibits the voids to fill asymptotically all the visible 
universe, namely $f_{v\infty}< 1$. Instead, an expected small fraction of space is asymptotically filled
by walls. The amount of such a fraction cannot be deduced by an asymptotic study 
of the equations and can be obtained only by
integrating the field equations (a numerical integration will be the
subject of a separate paper).
The expansion of the voids is different with respect to the FB model and could be in 
principle detectable. In fact, the role of the cosmological constant is to make the asymptotic
behaviour of $a(t), a_w(t), a_v(t)$ the same, contrary to the 
FB model, where the late time limit of $a_v(t)/a_w(t)$ is diverging. In particular, this 
phenomenon is visible in the
behaviour of $I_w$: this approaches the value $1$ both at early and late times. 
The behaviour of $I_w$
is also interesting when analyzed in terms of the $\Lambda$FB model with clock effects. 
In fact, we can add clock effects of \cite{12} in our model.  
In this case, the field equations
for $f_v, a(t), a_w, a_v(t)$ remain unchanged, while the parameter $I_w$ becomes 
a phenomenological lapse function. As a consequence, also in the 
presence of $\Lambda$, clock effects are possible: time delay effects can coexist with a 
non-vanishing cosmological constant and
we can also have cosmic clock effects which are interpreted in terms of quasi-local gravitational 
energy in presence of the mysterious $\Lambda$. A broadly uniform Hubble flow
can thus be obtained with smaller clock effects than in \cite{10,12}
and a non-zero cosmological constant. This reasoning shows that so called
Sandage-de Vaucouleurs paradox can also be solved with  $\Lambda\neq 0$.

As a further consideration, within our model we deduce a rather interesting formula
relating all the cosmological parameters of the $\Lambda$FB model at a given present time.
A remarkable consequence of this formula is the existence of a minimum actual value for
${\Omega}_{\Lambda}$ (at least for positive curvatures) 
expressed in terms of the age of the universe and the actual 
density parameters ${\Omega}_{k}$ and ${\Omega}_{m}$. 

As a final consideration, note that our model predicts the way in which voids 
expand in presence of $\Lambda$. This behaviour could be object of a cosmological test.
In fact, methods to measure the expansion rate of voids are in preparation (see for example
\cite{gg}). When these methods will be available, they will represent a way to distinguish between
the inhomogeneous models present in the literature.

\section*{Acknowledgements}
I would like to thank Alessandra D'Angelo for useful suggestions.

\section*{Appendix}
In our model, the observer is located within walls with zero curvature. 
The metric within the walls can be written as
\begin{eqnarray}
ds_{w}^2&=&-dt^2+a_{w}^2\left[d{\eta}_w^2+{\eta}_w^2d{\Omega}^2\right]\label{25}\\
        &=&-dt^2+\frac{{(1-f_v)}^{\frac{2}{3}}a^2}{f_{wi}^{\frac{2}{3}}}
          \left[d{\eta}_w^2+{\eta}_w^2d{\Omega}^2\right]\nonumber
\end{eqnarray}
The hyperbolic voids metric is:
\begin{eqnarray}
ds_{v}^2&=&-dt^2+a_{v}^2\left[d{\eta}_v^2+{\sinh}^2({\eta}_v) d{\Omega}^2\right]\label{26}\\
        &=&-dt^2+\frac{{f_v}^{\frac{2}{3}}a^2}{f_{vi}^{\frac{2}{3}}}
          \left[d{\eta}_v^2+{\sinh}^2({\eta}_v) d{\Omega}^2\right].\nonumber
\end{eqnarray}
It should be noticed that the scale factors $a_w,a_v$ are not the ones 
obtained by solving the exact Einstein equations but instead they are obtained by
solving the Buchert equations (\ref{23}) and (\ref{24}) with a non-vanishing
backreaction. Concerning the metric at the scale of homogeneity with the scale factor 
$a(t)$ given by (\ref{13}) we have:
\begin{equation}
ds^2=-dt^2+a^2d{\eta}^2+A(t,\eta)d{\Omega}^2,
\label{27}
\end{equation}
where $A(t,\eta)$ is an area function satisfying 
$4\pi\int_0^{{\eta}_{\mathcal{H}}}A d\eta=a^2V_i({\eta}_{\mathcal{H}})$,
where ${\eta}_{\mathcal{H}}$ is the particle horizon radius.
Note that also the metric (\ref{27}) is not an exact solution of Einstein equations,
i.e. it is not a LTB metric,
but rather it is obtained by solving the full Buchert equations (for more
on the interpretation of (\ref{27}) see \cite{12}).
The main astrophysical observations are by means of photons propagating
along null geodesics.
In the Buchert average scheme we do not average along the light cone.
Nevertheless, we assume that one can describe the light cone with the
averaged geometry
(\ref{27}). In practice, rather than averaging a bundle of null geodesics,
we assume that the light cone is described by means of the fiducial averaged geometry
(\ref{27}). Consequently, an observer located within walls, must relate his geometry
given by (\ref{25}) in term of the Friedmann fiducial metric (\ref{27}) along the past null
cone, i.e. the metric of the observer must be dressed (see \cite{7,8}) by means of
the  Friedmann bias given by (\ref{27}). 
In contrast to many approaches to the Buchert equations which ignore the issue of the dressing
of cosmological parameters and simply relate the volume average scale factor to the 
observed redshift \cite{z6,z7,z8}, we will adopt the matching procedure of Wiltshire \cite{12}
and match the radial null sections of
(\ref{25}) and (\ref{27}), with the phenomenological lapse function set to unity in our case. 
We obtain:
\begin{equation}
d{\eta}_w=\frac{f_{wi}^{\frac{1}{3}}d\eta}{{(1-f_v)}^{\frac{1}{3}}}.
\label{28}
\end{equation}
Thanks to (\ref{28}), the wall geometry (\ref{25}) becomes:
\begin{equation}
ds_w^2=-dt^2+a^2\left[d{\eta}^2+{(1-f_v)}^{\frac{2}{3}}
f_{wi}^{-\frac{2}{3}}{\eta}_w^2 d{\Omega}^2\right].
\label{29}
\end{equation}
The past null cone equation in (\ref{29}) is given by:
\begin{equation}
\eta=\int_{t}^{t_0}\frac{dt}{a}.
\label{30}
\end{equation}
The distance-redshift relation $d_L(z)$ in terms of the redshift $z$ is given by:
\begin{eqnarray}
& &d_L(z)=a_0(1+z){\overline{\eta}}_w,\label{31}\\
& &{\overline{\eta}}_w={(1-f_v)}^{\frac{1}{3}}
\int_t^{t_0}\frac{dt}{{(1-f_v)}^{\frac{1}{3}}a}.\nonumber
\end{eqnarray}
The functions $f_v,a$ are obviously obtained by solving the Buchert equations.

\end{document}